\documentclass{elsart}
\usepackage{natbib}
\input psfig.sty
\begin{document}
\runauthor{Dirk Grupe}
\begin{frontmatter}
\title{Statistical Properties of Narrow Line Seyfert 1 galaxies}
\author[garching]{Dirk Grupe\thanksref{email}}

\address[garching]{Max-Planck-Institut f\"ur extraterrestrische Physik,
Giessenbachstr., D-85748 Garching, Germany}
\thanks[email]{E-mail: dgrupe@mpe.mpg.de}
\begin{abstract}
The number of publications considering Narrow-Line Seyfert 1 galaxies
has increased dramatically in recent
years. Especially after the launch of the X-ray missions ROSAT and ASCA, 
Narrow-Line Seyfert 1s have become very popular. 
In these proceedings I will give an overview of how
they are distributed over the electromagnetic spectrum.
I will describe what we know 
about them at radio, infrared, optical, and X-ray bands, and  
how they differ and how they are similar to Broad-Line Seyfert 1s. 
Finally I will introduce a method
to find them with high probability. 
\end{abstract}
\end{frontmatter}

\section{Introduction}
\begin{figure}
\psfig{file=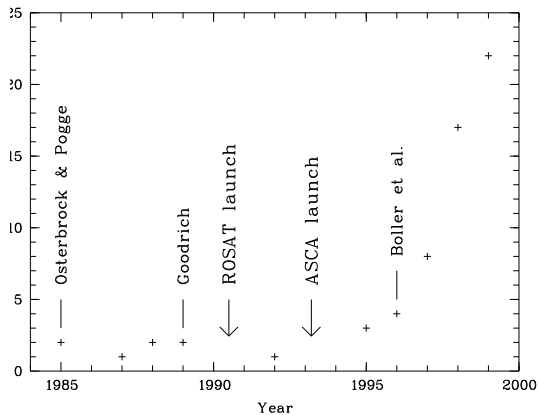,width=12.0cm,bbllx=3.0cm,bblly=1.0cm,bburx=28.0cm,bbury=19.5cm,clip=} 
\caption{\label{distr} Development of the number of publications per year that
contain the word 'Narrow Line Seyfert 1' in their title. The arrows mark the
launches of ROSAT and ASCA. Important publications about NLS1 are marked as well.
}
\end{figure}

Looking back in time, Narrow-Line Seyfert 1 galaxies (NLS1) have become one of
the most popular issues in astronomy 
in recent years. Between the definition of NLS1
by Osterbrock \& Pogge in 1985 and the launch of ROSAT in 1990, only a couple of
papers on NLS1 had been published. The launches of ROSAT and ASCA seem to be
the break points. After these launches in 1990 and 1993, respectively,
 the number of
publications that deal with NLS1 has increased dramatically.
In Figure \ref{distr} the number of
publications is shown that contain the term `Narrow Line Seyfert 1' in their
title, derived from the CDS/ADS abstract server. 
However, this is only the tip of the iceberg. The real number of
publications is much higher. NLS1 have become an important class of objects in
astrophysics. 

\section{A brief history of NLS1}
Narrow Line Seyfert 1 galaxies were defined by Osterbrock \& Pogge in
1985 (and Goodrich 1989)
by their optical properties (see also R.W. Pogge's article in these
proceedings). Their definition was that NLS1 are those AGN which have
FWHM(H$\beta$) $<$ 2000 $\rm km~s^{-1}$ and [OIII]/H$\beta$ flux ratio $<$ 3.0.
In this way they were able to distinguish these objects from Seyfert 1s and
Seyfert 2s (the H$\beta$ width for S1 and the [OIII]/H$\beta$ ratio for S2).
At this point it was not obvious how important NLS1 would become in the future.
NLS1 were rare objects and D. Osterbrock stated that only about 15\% of all
Seyfert 1 and 1.5s are NLS1 (Osterbrock, 1987). It was S. Stephens in 1989
who suggested from her studies on X-ray selected AGN, that X-ray selection would
be a good method to pick out these objects. This statement
was even more surprising because in
the X-ray study of Piccinotti et al. (1982) none of their sources fullfiled the
criteria for NLS1.  Boroson \& Green (1992) used the technique of Principal 
Component
Analysis (PCA) to search for an underlying component that governs the observed
parameters in AGN. They used 87 optically selected AGN and found an 
Eigenvector 1 (EV 1)
that combined parameters like the line width, [OIII]/H$\beta$, the FeII
strength, and the H$\beta$ asymetry.  A study of ultrasoft
EINSTEIN-selected AGN was presented by Puchnarewicz et al. (1992).
They found that objects with steep X-ray spectra have smaller
H$\beta$ widths than those with flatter X-ray spectra. Similar results were
found from ROSAT observations. Boller et al. (1996) used pointed ROSAT PSPC
observations of optically selected NLS1, and showed a clear correlation between
the width of H$\beta$ and the soft X-ray slope $\alpha_{\rm X}$ (see also Thomas
Boller's contribution in these prceedings).
Coming from the
other side of the electromagnetic spectrum, Grupe (1996) and Grupe et al. (1998a,
1999a) found that about 40\% of their 76 soft X-ray selected ROSAT AGN are NLS1.

To conclude, NLS1 have become very important in X-ray surveys. In the next
section we will have a look at several important X-ray samples and see how NLS1
appear in them.  

\section{\label{x-sample} X-ray selected AGN samples}

The first sample that showed that NLS1 appear enhanced in X-ray selected samples
was the EINSTEIN AGN sample of Stephens (1989). She found that 10 of her 42
Seyfert 1s were NLS1 (=25\%). This was more than expected
(see Osterbrock 1987) based on optically selected samples. 
Puchnarewicz et al. (1992) presented a sample of 52
serendipitously found EINSTEIN AGN. Out of their 17
Seyfert 1s, 9 turned out to be NLS1.
They also noticed the relation between the width of H$\beta$ and the X-ray
spectral slope. 

With the ROSAT All-Sky Survey (RASS, Voges 1993) it became
possible to study big
samples of AGN. Bade et al. (1995) presented a cross correlation between the
RASS and the Hamburg Quasar Survey and found 225 AGN in total, of which 55
fulfilled they criteria to be NLS1 (=25\%). Grupe et al. (1998a, 1999a) used 
bright soft X-ray selected AGN (for selection criteria see Thomas et al. 1998)
based on the ROSAT Bright Source Catalogue (Voges et al. 1998). 
The result was that out of the 76 objects of the sample, 25 were NLS1 using the
strict criteria, and 32 if the criteria are more relaxed. This means
that about 40\% of the objects in the soft X-ray selected sample are NLS1.
Another sample of soft X-ray AGN is the one of Edelson et al. (1999), who studied
the AGN discovered by the ROSAT Wide Field Camera. Of the 19 sources in their
sample, about 40\% are NLS1. 

So far it looks like using X-ray surveys can increase
the number of NLS1. However, is this true for all X-ray selected AGN samples?
Fischer et al. (1998) presented bright hard X-ray selected
AGN (sort of the continuation of the Grupe et al. sample towards harder X-ray
spectra) and found only 4 out of 31 Seyfert 1s to be NLS1. Puchnarewicz et al. (1996)
studied 85 AGN of the ROSAT International X-ray/Optical Survey (RIXOS, Mittaz et
al. 1999, Mason et al. 2000) 
that uses ROSAT pointings. 19 of these AGN turned out to be NLS1 
(=22\%).
Lehmann et al. (2000 and these proceedings) presented identifications of the 
ROSAT Deep Survey observations of the Lockman Hole objects (Hasinger et al.
1998, see also these proceedings) 
and found only 1 out of 33 Seyfert 1 to be a NLS1. A similar result was
found by Appenzeller et al (1998) and Krauter et al. (1999) who identified the
objects in 6 selected ROSAT fields at Dec = $+ 9^{\circ}$. Only 3 out of
149 Seyfert 1s
were NLS1. It is also interesting to mention that in the bright soft ROSAT survey
of Schwope et al. (2000), of the almost 700
AGN, only 4\% are NLS1.

To summarize this section, Stephens was right in saying
that using X-ray surveys can
increase the number of NLS1. We must be careful though, as
 not all X-ray surveys
guarantee a
number of them. How we can find a method that gives a good chance
of finding 
NLS1, we will see later.

\section{Properties of NLS1 over the electromagnetic spectrum}

\subsection{The famous FWHM(H$\beta$) vs. $\alpha_{\rm X}$ relation}

Puchnarewicz et al. (1992) suggested it, and Boller et al. (1996) finally 
showed it, i.e. the
relation between the width of the broad H$\beta$ line and the X-ray spectral
slope $\alpha_{\rm X}$ (see Figure 8 in Boller, Brandt \& Fink (1996) and
these proccedings).
Objects with narrow
H$\beta$ emission from the Broad Line Region (BLR) tend to have steeper X-ray
spectra. Only NLS1 show very steep X-ray spectra.
Grupe et al. (1999a) found that this correlation was more pronounced among
high-luminosity AGN. Also Laor et al. (1997) found a clear correlation betwen
these properties among the objects of their complete sample of 23 optically
selected quasars. The trend that NLS1 have steeper X-ray spectra
does not only appear in the soft X-ray band it also
continues to harder energies as shown by Brandt et al. (1997, see their Figure 1)
for ASCA spectra of Seyfert 1s. K. Leighly (1999a,b) found that NLS1 have
significantly steeper ASCA spectra than Broad Line S1 (see also her contribution
to these proceedings).  

\subsection{Other optical properties and PCA}
It has been found in several samples that NLS1 have stronger FeII emission
than BLS1. They also tend to have weaker emission from NLR lines. Boroson \&
Green discovered the anti-correlation between those two properties. 
By using Principal Component Ananlysis (PCA), they found
that several optical properties of AGN are linked and governed by one fundamental
parameter. The nature of this fundamental parameter  is
still unclear. It may be the mass of the central black hole, the
accretion rate or even the age of the AGN 
(see Grupe 1996, Grupe et al. 1999a). NLS1 lie at the
extreme end of the Eigenvector 1 relation of AGN. 

\subsection{Radio observations of NLS1}

So far only one radio survey on NLS1 has been published: Ulvestad et al. (1995)
used the NLS1 sample of Goodrich (1989) obverved with the VLA. 
They found that
NLS1 have typical radio luminosities for Seyfert 1 galaxies. 
The radio emission is emitted from the core. 

A new study of more than 20 NLS1
has been performed by Moran et al. (in prep, see also these
proceedings), using the IRAS vs. RASS sample of Boller et al. (1992,
1997). 

Boroson \& Green (1992) stated in their study that `the FeII emission is weak in
radio-loud objects'. Therefore it was a surprise that some NLS1 have turned out
to be radio-loud. Three sources have already been published: PKS 0558--504
(Remillard et al. 1991), RGB 0044+193 (Siebert et al. 1999), and RX
J0134.2--4258 (Grupe et al. 2000).  Several more radio-loud NLS1 have been found
in the Moran et al. sample and in the FIRST survey, and remain to be published
(E. Moran, private communications). 

\subsection{Infrared Studies}
As in the radio band, the infrared lacks studies of big NLS1 samples. Only a
couple of sources have been studied at IR wavelengths. Pretty early on,
just after
the Osterbrock \& Pogge NLS1 paper (1985), Halpern \& Oke presented a study
of
three NLS1 (Mkn 507, 5C 3.100, I Zw 1), and found those sources to be more
luminous in the infrared than BLS1. A similar result was found by Moran et al.
(1996). On the other hand, Poletta \& Courvoisier (1999, see also these
proceedings) used ISOPHOT observations of 16 NLS1, and found that they have
similar luminosities to BLS1. Also Murayama et al. (2000) found from IRAS
observations at 25$\mu$m that the luminosities of NLS1, and BLS1 are similar at
this wavelength.

\subsection{Optical polarimetry} 

One of the first explanations of why NLS1 do not show very broad BLR
emission lines was that the high velocity component of
the BLR might be hidden, as has been proposed to explain the difference
between S1 and S2. In this case, polarization by scattered light would be
expected. R. Goodrich (1989) presented a study of 18 NLS1 and found that 3 of 
his
sources do have significant degrees of polarization ($>$ 2\%). The 
wavelength dependence of the degree of polarization means, that  part 
of the emission is
scattered. He also found that the permitted lines are nore, and the forbidden
lines less polarized than the continuum, suggesting that the scattering
material is somewhere between the BLR and NLR. 

Grupe et al. (1998b) performed an optical polarimetry survey of all objects in
their bright soft X-ray selected ROSAT AGN sample, and found that most of the
sources turned out to be either unpolarized or only polarized at very low
levels. This result
suggests a direct unabsorbed view of the central engine of the
AGN. Two NLS1 of that sample, however, turned out to have wavelength dependent
polarization of $>$ 4\% in V. Both are IRAS galaxies, IRAS 13349+2438 (Wills
et al. 1992) and IRAS 12397+3333 (Grupe et al. 1999c, Wills et al. in prep). 
These two objects both show strong warm absorber features in their ROSAT and ASCA
spectra (see Brandt et al 1996, 1997 for IRAS 13349+2438, Grupe et al. 1999c and
Wills et al. in prep. for IRAS 12397+3333). Actually, Leighly et al. (1997) have
shown that there is a trend towards higher degrees of polarization if warm
absorbers are present. 

\subsection{Variability studies}

\subsubsection{Long-term variability}
Studies of optical variability are common for several BLS1, e.g. NGC 5548.
However, studies of NLS1 are rare. Giannuzzo \& Stirpe (1996) and Giannuzzo et
al. (1998) presented a study of the optical variability within
a sample of NLS1 and
found similar variability as BLS1. H.R. Miller found in a study of 5 NLS1 that
only one source has shown significant variability in the optical  band
(see his
article in these proceedings).

Variability is normal for AGN in the X-ray band and factors of 2-3 in the
X-ray band are quite common. 
K. Leighly (1999a,b) has
presented a comprehensive study of 23 NLS1 with ASCA. She found that the excess
variance that describes the variability in the overall light curves, is bigger in
NLS1, compared with BLS1 (see Figure 3 in Leighly 1999a, and also these
proceedings).

\subsubsection{Giant and rapid X-ray variability}

Some NLS1 have shown persistent giant and rapid X-ray variability. In the most
extreme case, IRAS 13224--3809 (Boller et al. 1993, 1997), flux changes of
factors of 57 in 2
days have been observed. The shortest doubling time scale observed in this
object was only 800s, suggesting that the X-ray emission is coming from the
innermost part of the AGN engine. IRAS 13224--3809 is one of the few NLS1 in
which optical variability has been seen (see H.R. Miller contribution in these
proceedings). Another X-ray rapidly variable source is PHL 1092 (Forster \&
Halpern, 1996). For this object, 
Brandt et al. (1999) suggested relativistic beaming as
the explanation for the rapid variability.

\subsubsection{X-ray transience}
Since the RASS, a new class of AGN has been established: X-ray transient AGN.
These objects appear to be X-ray bright only once and then get fainter or even
vanish when observed years later. 
This phenomenon has been observed in NLS1, such as WPVS007
(Grupe et al. 1995b) or RX J0134.2+4258 (Grupe et al. 2000, and these
proceedings).  However, it is not
specific to NLS1. X-ray outbursts have been observed in rather low-active AGN
(like the Seyfert 2
IC 3599, Brandt et al. 1995, Grupe et al. 1995a) or even non-active
galaxies, e.g. RX J1624.9+7554 (Grupe et al. 1999b) or RX J1246.6--1119 (Komossa
\& Greiner (1999)).

\section{How to find NLS1}
We have seen that the number of NLS1 found over the last couple of years has 
increased
dramatically. Is there a method to find them? Can we go `fishing' for them and
have a good chance that the right `fish' sticks bites. Stephens (1989)
suggested using X-ray surveys, and a couple of samples support this suggestion
(see Section \ref{x-sample}).  However, in several X-ray
samples only a small
number of NLS1 have been found. The criterion of X-ray survey alone can not be
enough to find NLS1. The conclusion we can draw from looking at all those
X-ray samples is that NLS1 can be selected from objects with steep soft X-ray
spectra. In addition, to get a higher probability, the $f_{\rm X}/f_{\rm
opt}$ has als to be taken into account (see Figure 2 in Beuermann et al. 1998). AGN
group around log $f_{\rm X}/f_{\rm opt}~\approx~0.5$. Referring to this diagram,
my experience is if you take objects around this log $f_{\rm X}/f_{\rm opt}$
value and with HR1 $<$ --0.5,
you will have a chance of about 75\% of finding NLS1. 

To conclude, Sally Stephens was right in suggesting X-ray samples to look for
NLS1, but the objects have to be soft X-ray sources. This also explains why in
the Piccinotti et al. (1982) sample, no NLS1 were found. All those objects had
hard X-ray spectra.

\section{Summary}
NLS1 are AGN with extreme properties. They show the steepest X-ray spectra, the
strongest FeII emission and the lowest emission from NLR lines. They are more
variable in X-rays than BLS1 and their soft X-ray excess is stronger. Soft X-ray 
selected NLS1 do
not show significant cold absorption and polarization. However, optically
selected NLS1 do (see Goodrich 1989).

One question still remains: Are NLS1 their own class or are they just a subclass
of Seyfert 1s? Well, their properties are extreme and NLS1 are the objects with the
steepest X-ray spectra and the most pronounced soft X-ray excess.
However, there
is no distinct physical boundary between NLS1 and BLS1. My personal point of
view  is, that they are just the continuation of Seyfert 1s towards extreme properties. They
are at the end of the Boroson \& Green (1992) Eigenvector 1 relation. NLS1 are
probably AGN that accrete at Eddington accretion rates, or if Eigenvector 1
represents
the age of AGN, they can be considered to be very young objects (Grupe 1996, see
also S. Mathur's contribution in these proceedings, and her recent article
(2000)).

One final remark: The number of NLS1 that have been found 
has indeed increased dramatically  in the last decade. 
Nevertheless, NLS1 existed of course before Osterbrock \& Pogge (1985)
defined them. It took years to make the term `Narrow-Line Seyfert 1' popular.
Many objects which we would consider today to be NLS1 were not named so in
those days. On the other hand today, objects are offen called NLS1 which do not
belong in this category.

{\em Acknowledgments:} 

I would like to thank Drs. Bev Wills and Thomas Boller for carefully reading the
manuscript and their suggestions and comments to this article.

\end{document}